\documentclass[12pt]{elsart}
\usepackage{epsfig}

\def\um{\ifmmode {\mathrm{\mu m}}\else
                  \textrm{$\mu$m }\fi}%
\def\GeV{\ifmmode {\mathrm{\ Ge\kern -0.1em V}}\else
                   \textrm{Ge\kern -0.1em V}\fi}%
\def\MeV{\ifmmode {\mathrm{\ Me\kern -0.1em V}}\else
                   \textrm{Me\kern -0.1em V}\fi}%
\def\keV{\ifmmode {\mathrm{\ ke\kern -0.1em V}}\else
                   \textrm{ke\kern -0.1em V}\fi}%
\def\eV{\ifmmode  {\mathrm{\ e\kern -0.1em V}}\else
                   \textrm{e\kern -0.1em V}\fi}%
\def\uW{\ifmmode  {\mathrm{\mu  W}}\else
                   \textrm{$\mu$W}\fi}%
\def\thefootnote{\fnsymbol{footnote}}

\begin{document}

\begin{frontmatter}
\date{01 Mar 2006, v2.c}


\title{ Data production of a large Linux PC Farm for 
        the CDF experiment }

\centering{\large 
  J.~Antos$^{a}$,
  M.~Babik$^{a}$,
  A.W.~Chan$^b$, 
  Y.C.~Chen$^b$, 
  S. Hou$^b$ \footnote{E-mail: suen@fnal.gov}, 
  T.L.~Hsieh$^b$,
  R.~Lysak$^b$,
  I.V.~Mandrichenko$^c$, 
  M.~Siket$^{b,d}$,
  J.~Syu$^c$,
  P.K.~Teng$^b$,
  S.C.~Timm$^c$,
  S.A.~Wolbers$^c$,
  P.~Yeh$^b$
}

\address{
  $^a$ Institute of Experimental Physics, Slovak Academy of Sciences, 
       Slovak Republic  \\
  $^b$ Institute of Physics, Academia Sinica, Nankang, Taipei, Taiwan \\
  $^c$ Fermi National Accelerator Laboratory, Batavia, IL, USA \\
  $^d$ Comenius University, Slovak Republic
}

\begin{abstract}

The data production farm for the CDF experiment is designed and 
constructed to meet the needs of the Run II data collection at 
a maximum rate of 20 MByte/sec during the run. 
The system is composed of a large cluster of personal computers (PCs)
with a high-speed network interconnect and a custom design control 
system for the flow of data and the scheduling of tasks on this PC farm.
The farm explores and exploits advances in computing and communication 
technology.  The data processing has achieved a stable production
rate of approximately 2 TByte per day.
The software and hardware of the CDF production farms has
been successful in providing large computing and data throughput 
capacity to the experiment. 

\end{abstract}
\end{frontmatter}

\section{Introduction}

High-Energy Physics has advanced over the years by the use of 
higher energy and higher intensity particle beams, more capable 
detectors and larger volumes of data.  
The Tevatron Collider at Fermilab is used to study fundamental 
properties of matter by colliding protons and anti-protons at very high
energy.
The Fermilab Tevatron Run II project has increased the intensity 
and energy of the proton and anti-proton beams.
The Collider Detector at
Fermilab (CDF) detector is a large general purpose cylindrical detector
used to measure charged and neutral particles that result from the
proton-anti-proton collision.
The CDF detector has been upgraded to take advantage of the 
improvements in the accelerator~\cite{CDF2}.  
Computing systems were also upgraded for processing larger 
volumes of data collected in Run II.

The type of computing required for CDF data production can be 
characterized as loosely-coupled parallel processing~\cite{wolb}.
The data consists of a group of ``events", where each event is the 
result of a collision of a proton and an anti-proton.
A hardware and software trigger system is used to
store and save as many of the most interesting collisions as possible.
Each event is independent in the sense that it can be processed through
the offline code without use of information from any other event.
Events of a similar type are collected into files of a data stream.
Data is logged in parallel to eight data streams for
final storage into a mass storage system.

Each file is processed through an event reconstruction program that 
transforms digitized electronic signals from the CDF sub-detectors 
into information that can be used for physics analysis. 
The quantities calculated include particle trajectories and momentum, 
vertex position, energy deposition, and particle identities.

The CDF production farm is a collection of dual CPU PCs running Linux,
interconnected with 100 Mbit and gigabit ethernet.
This farm is used to perform compute and network intensive tasks 
in a cost-effective manner and is an early model for such computing.
Historically, Fermilab has used clusters of processors to provide large
computing power with dedicated processors (Motorola
68030)~\cite{gaines} or commercial UNIX workstations \cite{rinaldo}.
Commodity personal computers replaced UNIX
workstations in the late 1990s.
The challenge in building and operating such a system is in 
managing the large flow of data through the computing units.

This paper will describe the hardware integration and software for
operation of the CDF production farm. 
The first section will describe the requirements and design goals 
of the system.  Next, the design of the farm, hardware and software,
will be given.  The software system will be described in the next 
section. Next, the performance and experiences with the system, 
including prototypes, will be described. 
Finally, conclusions and general directions for the future are given.

\section{System Requirements}

To achieve the physics goals of the CDF experiment at the Fermilab
Tevatron, the production computing system is required to process the
data collected by the experiment in a timely fashion.
The CDF production farm is required to reconstruct the raw data with 
only a short delay that allows for the determination and availability 
of calibrations or other necessary inputs to the production executable.
In addition the farm is expected to reprocess 
data and to process special data.

To accomplish rapid data processing through the farms, adequate 
capacity in network and CPU is required.
In 2001 through 2004 the CDF experiment collects a maximum 
of 75 events/second at a peak throughput of 20 MByte/sec.
The event processing requires 2-5 CPU seconds on a Pentium III 1 GHz PC.
The exact number depends on the type of event, the version of the 
reconstruction code, and the environment of the collision. 
These numbers lead to requirements of the equivalent of 190-375 
Pentium III 1 GHz CPUs, assuming 100\% utilization of the CPUs. 

The output of event reconstruction is split into many physics data-sets.
The splitting operation is required to place similar physics data 
together on disk or tape files, allowing faster and more efficient 
physics analysis.
The output event size is currently approximately the same as the input.
Each event is written 1.2 times on average because some events 
are written to more than one output data set.
Therefore the system output capacity is also required to be 
approximately 20 MByte/sec.

In addition to providing sufficient data flow and CPU capacity for
processing of data, the production farm operation is required to be
easily manageable, fault-tolerant, scalable, with good monitoring and 
diagnostics.
Hardware and software options were explored to meet the
requirements for the system.  These include large 
symmetric multiprocessing (SMP) systems,
commercial UNIX workstations, alternative network configurations.
Prototype systems were built and tested before the final design was
chosen and production systems built.

\begin{figure}[b!]
  \vspace{-.5cm}
  \centering
  \epsfig{file=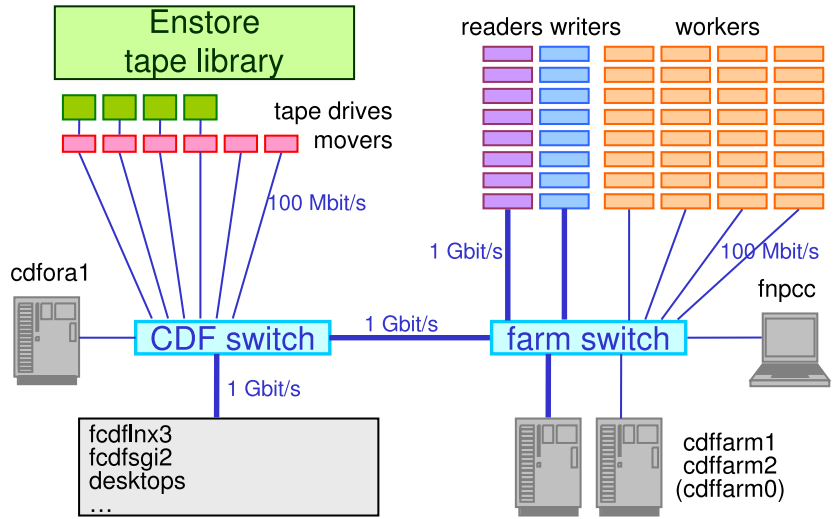,width=0.85\linewidth}
  \caption{CDF production farm architecture.
  \label{fig:farm} }
  \vspace{-.5cm}
\end{figure}

\section{Architecture and data flow }

The CDF data production farm is constructed using cost-effective 
dual CPU PC's.
The farm consists of a large number
of PCs that run the CPU-intensive codes (workers), PCs that buffer data
into and out of the farm (readers and writers) and PCs
providing various services (servers).
The hardware architecture of the CDF production farm is shown in 
Fig.~\ref{fig:farm}.
It has two server nodes {\sf cdffarm1} and {\sf cdffarm2}.
{\sf cdffarm1} is a SGI O2000 machine 
that host a batch submission system and a database server.
{\sf cdffarm2} is a dual Pentium server running control daemons 
for resource management and job submission.
These two servers have recently been replaced by a Dell 6650 machine ({\sf cdffarm0}).
Monitoring and control interfaces for farm operation includes a 
java server to the control daemons and and a web server for monitoring.
The disk space is a ``dfarm'' file system \cite{dfarm},
which is a distributed logical file system using a collection of
IDE hard-disks of all dual Pentium nodes.
The dfarm server is hosted on {\sf cdffarm1}.

The job scheduling on the production farm is controlled by a batch 
management system called FBSNG developed by the Computing Division 
at Fermilab \cite{FBS}.
The CDF Data Handling group has well-defined interfaces and operation
\cite{data_handl} to provide input data for the farm and to 
write output to a mass storage system (Enstore) \cite{Enstore}. 

\begin{table}[b!]
  \begin{center}
  \begin{tabular}{|c|c|c|l|} \hline
   Year & Total & Type & P-III eq. GHz    \\ \hline
   1998 &  1 &  SGI O2000     & (cdffarm1)  \\ \hline
   1998 &  1 &  PC server     & (cdffarm2)  \\  \hline
   1999 & 50 &  Pentium III/0.50 duals & (retired)     \\ \hline
   2000 & 23 &  Pentium III/0.80 duals & (retired)     \\ \hline
   2001 & 64 &  Pentium III/1.00 duals & 128         \\ \hline
   2002 & 32 &  Pentium III/1.26 duals & 81           \\ \hline
   2002 & 32 & AMD/1.67 duals & 107       \\ \hline
   2003 & 64 &  Pentium 4 Xeon/2.67 duals  & 253$^*$   \\ \hline
  \end{tabular}

  \caption{Farm nodes added over the years
     ($^*$ Xeon CPU is scaled by 1.35 to the equivalent Pentium III CPU 
     performance for CDF data reconstruction).
     The total in use in the summer of 2004 is 192 nodes (570 GHz).
  \label{tab:workers} }
  \end{center}
\end{table}

The dual Pentium nodes purchased over the years are
listed in Table~\ref{tab:workers}.
Old nodes were replaced after three years service.
At its peak in mid-2004, there were 192 nodes in service
including 64 dual Pentium 4 2.6 GHz machines added in the spring 2004.
The dfarm capacity of the collected worker hard-disks was as
large as 23 TByte including three
file-servers each having 2 TByte.
The IDE hard-disk size varies from 40 to 250 GByte.

The input and output (I/O) nodes are configured to match 
the data through-put rate.
A total of 16 nodes equipped with optical giga-links
are configured with the {\sf pnfs} file system \cite{pnfs}
for access to the Enstore storage.
A 48 port Cisco switch module was added recently to provide gigabit
Ethernet over copper switching.
Additional I/O nodes may be added if needed.
The number of workers can be scaled to as large a number as is required.
However, the total data through-put capacity to
Enstore storage is limited by the number of Enstore movers 
(tape-drives) available.

\begin{table}[b!]
  \begin{center}
  \begin{tabular}{|c|c|c|c|c|} \hline
   Stream  
   & data-sets & events/GByte & total event (\%) & total size (\%) \\
   \hline
   A & aphysr &  2720      &  3.8  & 7.7 \\
   B & bphysr &  5470      &  9.9  & 5.5 \\
   C & cphysr &  6770      &  9.2  & 7.5 \\
   D & dphysr &  2570      &  3.7  & 7.9 \\
   E & ephysr &  5930      &  17.0 & 15.7 \\
   G & gphysr &  6140      &  26.4 & 23.5 \\
   H & hphysr &  6050      &  19.6 & 17.7 \\
   J & jphysr &  5520      &  10.3 & 10.3 \\ \hline
  \end{tabular}
  \caption{Statistics of data streams of a typical run taken in 
     June 2004 containing all sub-detectors. 
     The raw data files are 1 GByte in size.
     Listed are the number of events per GByte,
     ratio of total events and total file size.
  \label{tab:rawdata} }
  \end{center}
  \vspace{-.5cm}
\end{table}

Raw data from the experiment is first written to tape in the 
Enstore mass storage system.
Raw data are streamed into eight data-sets listed in 
Table~\ref{tab:rawdata}.
These tapes are cataloged in the CDF Data File Catalog (DFC) \cite{DFC}
as a set of tables in an Oracle database 
(accessed via {\sf cdfora1} in Fig.~\ref{fig:farm}).
After the data is written to 
tape and properly cataloged, and once the necessary calibration 
constants exist, the data is available for reconstruction on the farms.

\begin{figure}[t!]
  \centering\epsfig{file=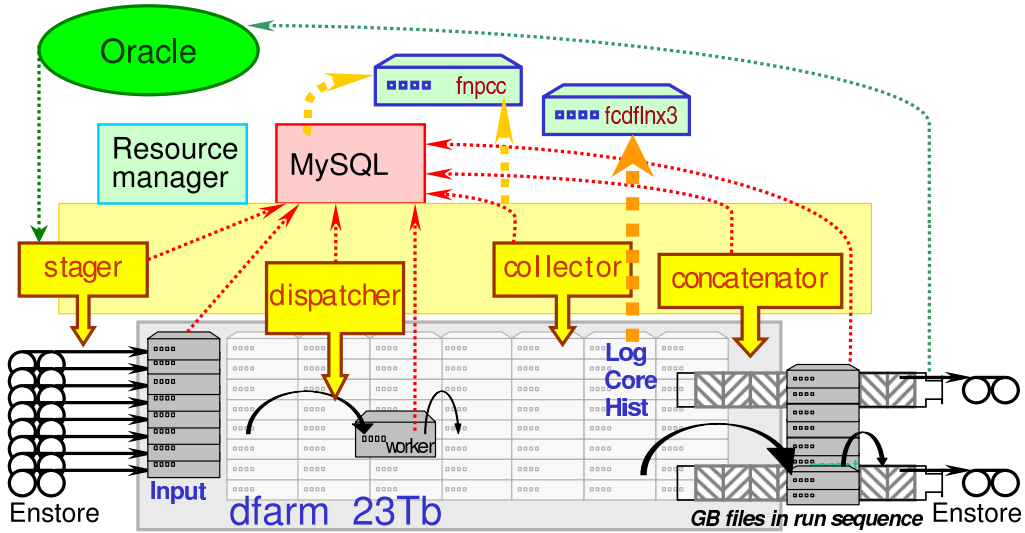,width=1.\linewidth}
  \caption{Flow control in the CDF production farm.
     Congestion is observed in concatenation waiting for lost files
     and in part caused by slow MySQL service.
  \label{fig:flow} }
  \vspace{.5cm}
\end{figure}

The production farm is logically a long pipeline with the constraint 
that files must be handled in order.
The input is fetched directly from Enstore tapes and 
the outputs are written to output tapes.
The data flow is illustrated in Fig.~\ref{fig:flow}
for the files moving through dfarm storage controlled by four
production daemons.  The daemons communicate with the
resource manager daemon and the internal database to schedule
job submission.
The internal database is a MySQL \cite{MySQL} system used for 
task control, file-tracking, and process and file history.
The DFC records are fetched at the beginning of staging input data.
Output files written to tapes are recorded in the DFC. 
Job log files and other logs and files are collected to the user 
accessible {\sf fcdflnx3} node. 
Operation status is monitored by a web server {\sf fnpcc}.

The operation daemons are configured specifically for 
production of a input ``data-set''.
For raw data, each data stream is a data-set.
The input files are sent to worker nodes for reconstruction.
Each worker node (dual-CPU) is configured to run two 
reconstruction jobs independently. An input file is 
approximately 1 GByte in size and is expected to run for about 
5 hours on a Pentium III 1 GHz machine.
The output is split into multiple files, with each file corresponding
to a data-set defined by the event type in the trigger system.
An event may satisfy several trigger patterns and is consequently 
written to multiple data-sets that are consistent with that event's 
triggers. 
Each data-set is a self-contained sample for physics analysis. 
The total number of output data-sets is 43 for the eight data streams 
used in the most recent trigger table.

\begin{figure}[t!]
  \centering\epsfig{file=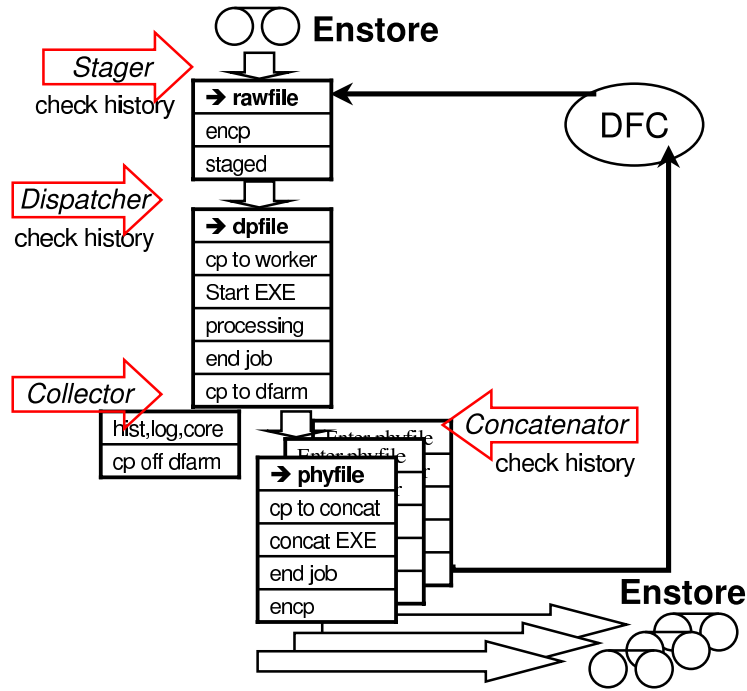,width=.75\linewidth}
  \vspace{-.5cm}
  \caption{Task control for a farmlet. Status is recorded for each
  input file in MySQL database.
  \label{fig:bookkeeping} }
\end{figure}

\section{CDF Farm Processing System (FPS)} 

\subsection{FPS daemons } 

The CDF Farm Processing System (FPS) is the software that
manages, controls and monitors the CDF farm.  
It has been designed to be flexible and allows configuration
for production of data-sets operated independently 
in parallel farmlets.
A farmlet contains a subset of the farm resources
specified for the input data-set, the executable 
and the output configuration for concatenation.
Since a farmlet is an independent entity, it is treated as such, 
that is, its execution is handled by its own daemons taking care of 
consecutive processing in production and its records are written 
in the internal database.  The task control by FPS for
a farmlet is illustrated in Fig.~\ref{fig:bookkeeping}.
The daemons of the FPS farmlets are :

\begin{itemize}
   \item {\bf Stager} is a daemon that is responsible for finding and 
   delivering data from tapes based on user selection for a set of 
   data files or run range in the data-set. 
   Jobs are typically submitted one ``file-set'' at a time.
   A file-set is a collection of files with a typical size of 10 GByte.
   The stager fetches DFC records for input and checks that
   proper calibration constants are available.
   The staging jobs are submitted to the input I/O nodes 
   and the file-sets are copied to their scratch area, 
   and afterward to dfarm.

   \item {\bf Dispatcher} submits jobs through the batch manager to 
   the worker nodes and controls their execution. 
   It looks for the staged input file, which is then
   copied into the worker scratch area.
   The binary tarball (an archive of files created with the Unix
   tar utility) containing the executable, complete libraries,
   and control parameter files are also copied. 
   This allows the reconstruction program to run locally on the 
   worker nodes and the output files, of various sizes from 5 MByte 
   to 1 GByte, are written locally.
   At the end of the job the output files are then copied back to dfarm.
   In case of abnormal system failure, job recovery is performed
   and the job is resubmitted.

   \item {\bf Collector} gathers any histogram files, log files and any 
   additional relevant files to a place where members of the 
   collaboration can easily access them for the need of validation or 
   monitoring purposes.

   \item {\bf Concatenator} writes the output data that is produced 
   to the selected device (typically the Enstore tape) in a timely 
   organized fashion. 
   It checks the internal database records for a list of files to be 
   concatenated into larger files with a target file size of 1 GByte.
   It performs a similar task as the dispatcher, with 
   concatenation jobs submitted to output nodes.
   The output nodes collect files corresponding to a file-set size
   ($\approx 10$ GByte) from dfarm to the local scratch area, 
   execute a merging program to read events in the input files
   in increasing order of run and section numbers.
   It has a single output truncated into 1 GByte files.
   These files are directly copied to tapes and DFC records are written.
\end{itemize}

Since all of the farmlets share the same sets of computers and
data storage of the farm, the resource management is
a vital function of FPS for distribution and prioritization 
of CPU and dfarm space among the farmlets. 
The additional daemons are:
\begin{itemize}
   \item {\bf Resource manager} controls and grants allocations for 
   network transfers, disk allocations, CPU and tape access based on a 
   sharing algorithm that grants resources to each individual farmlet 
   and shares resources based on priorities.  This management of 
   resources is needed in order to prevent congestion either on the 
   network or on the computers themselves and to use certain resources
   more effectively.
        
   \item {\bf Dfarm inventory manager} controls usage of the 
   distributed disk cache on the worker nodes that serves as a 
   front-end cache between the tape pool and the Farm.  

   \item {\bf Fstatus} is a daemon that checks periodically whether 
   all of the services that are needed for the proper functioning of 
   the CDF production farm are available and to check the status 
   of each computer in the farm. 
   Errors are recognized by this daemon and are reported 
   either to the internal database which can be viewed on the web or 
   through the user interfaces in real time. Errors can also be sent 
   directly to a pager with a copy to an e-mail address that is 
   registered as the primary recipient of these messages.
\end{itemize}

The FPS framework is primarily coded in python \cite{python}.
It runs on one of the server computers ({\sf cdffarm2}) and depends on 
the kernel services
provided by {\sf cdffarm1}, namely the FBSNG batch system, 
the FIPC (Farm Interprocess communication) between the daemons and 
dfarm server governing available disk space on the worker nodes.  
Daemons have many interfacing components that allow them to communicate
with the other needed parts of the offline architecture of the CDF 
experiment. 
Those include mainly the DFC (Data File Catalog) and the Calibration 
Database.

\subsection{Bookkeeping and internal database}

With hundreds of files being processed at the same time it is 
important to track the status of each file in the farm.
File-tracking is an important part of FPS and the bookkeeping is 
based on a MySQL database. 
The database stores information about each individual file, process 
and the history of earlier processing.
Three tables are implemented for each farmlet: for stage-in of input
files; reconstruction and output files; and the concatenation.
The processing steps tracked by the book-keeping and records in each 
table are illustrated in Fig.~\ref{fig:bookkeeping}. 
Once a file is successfully processed,
its records are copied over to the corresponding history tables.
The file status is used in order to control the flow of data and 
to make sure that files are not skipped or processed more than once. 
The MySQL database also includes detailed information about the 
status of each file at every point as it passes through the system.
This information is available through a web interface to the 
collaboration in real time.
This database server was designed to serve thousands of 
simultaneous connections. 
For our application it is a perfect match.

Emphasis is put on automatic error recovery and minimal human 
interaction in the course of processing.  With the help of information
that is stored in the internal database, the system is able in most 
cases to recover and return to the previously known state from 
which it can safely continue to operate.
The daemons checking the file history in the database are not 
instrumented to detect an abnormal failure for a job in process 
or a file lost to network or hardware problems.
The concatenator often has to wait for output file in order to 
combine files in order.
This bottleneck can be a serious problem and is a major consideration 
for relaxing strict ordering of files to improve overall 
system performance.

\subsection{User and Web Interface }

The FPS system status is shown in real time on a web page that gives
the status of data processing, flow of data, and other useful 
information about the farm and data processing.
The web page is hosted on a dual Pentium node 
({\sf fnpcc} on Fig.~\ref{fig:farm}) connected to the farm switch.
The web interface was coded in the PHP language \cite{php}
and RRDtool \cite{rrd} for efficient storage and display 
of time series plots.
The structural elements in the schema include
output from FPS modules, a parser layer that transforms data into a 
format suitable for RRDtool, a RRDtool cache that stores this data in 
a compact way, and finally the web access to RRD files and queries 
from MySQL for real time display of file-tracking information.

The java FPS control interface was designed for platform independent 
access to production farm control using an internet browser.
Information transfer
between the client and server over the network is done using IIOP 
(Internet Inter-ORB protocol) which is part of CORBA\cite{corba}.
It has proved to be stable, and there have been no problems with 
short term disconnections and reconnections. 
An XML processor\cite{xml} is used to generate and interpret the 
internal representation of data. Abstract internal representation 
of data is important to cope with changes in the FPS system.
A Java programming language, Java Web Start technology \cite{webs} 
was used for implementation of a platform independent client.

\section{Experience with Production }

\subsection{Early commissioning}

The CDF experiment collected data samples
in the Tevatron Run II commissioning run in October, 2000 
and the beginning of proton-antiproton collisions in April, 2001. 
These events were processed through the CDF production farms. 
The events collected during the commissioning run 
were processed through two versions of the reconstruction code.  
The early 2001 data, taken under various beam and detector conditions,
consist of about 7.6 million events and these were processed with one 
or two versions of the reconstruction code.  
This processing experience gave some confidence that the farm had 
the capacity to handle the volume of data coming from the detector 
and also uncovered many operational problems that had to be solved.

Beginning in June, 2001, both the Tevatron and the CDF detector ran 
well and began to provide significant samples for offline 
reconstruction. This early data was written in 4 streams and 
the output of the farms was split into 7 output data-sets. 
The CDF experiment wrote data at a peak rate of 20 MByte/sec, 
which met the design goal. The farms were able to reconstruct data at 
the same peak rate. The output systems of the farm were adjusted
to increase their capacity to handle the large output of the farms. 
More staging disk was added to provide a larger buffer 
and additional tape-drives were added.

Beginning in early 2002 the CDF detector and accelerator had reached a
point where data was being recorded in the 8 final data streams defined
for Run 2 and the output was split into the final physics data-sets
(approximately 50 different data-sets).  Data was processed as quickly 
as possible and was normally run through the farms within a few days of
having the final calibrations.  Approximately 500 million events were
collected and processed during this period.  Upgrades were made to the
farm with the addition of new nodes for processing as well as improved
I/O capability.  These improvements helped to maintain the processing
capability as well as to provide some capability to catch up when
calibrations were not ready in time or when the data taking rate was
high or when reprocessing was necessary.

A major reprocessing of all data collected from the beginning of 2002 
was begun in the fall of 2003 using the version of cdf code 5.1.1. 
The output of this processing was later
reprocessed with improved calibration (version 5.3.1)
of calorimetry and tracking leading to higher efficiencies.
The reprocessing was launched in March, 2004 and the production
farm operated at full capacity for a six week period.
The main characteristics and performance of the farm is described 
in the following sections.

\subsection{Data processing capacity}  

The CPU speed and data through-put rate are the factors that determine 
the data reconstruction capacity of the production farm.
The computing time required for an event depends on the 
event characteristics determined by the event trigger
in different data streams. 
In addition, the intensity of the proton and antiproton beams matters.
More intense beams lead to multiple events per beam
crossing which in turn lead to more CPU time per event.  
Inefficiency in utilizing CPU comes from the file transfer of
the executable and data files to and from the worker scratch area.

\begin{figure}[t!]
  \centering\epsfig{file=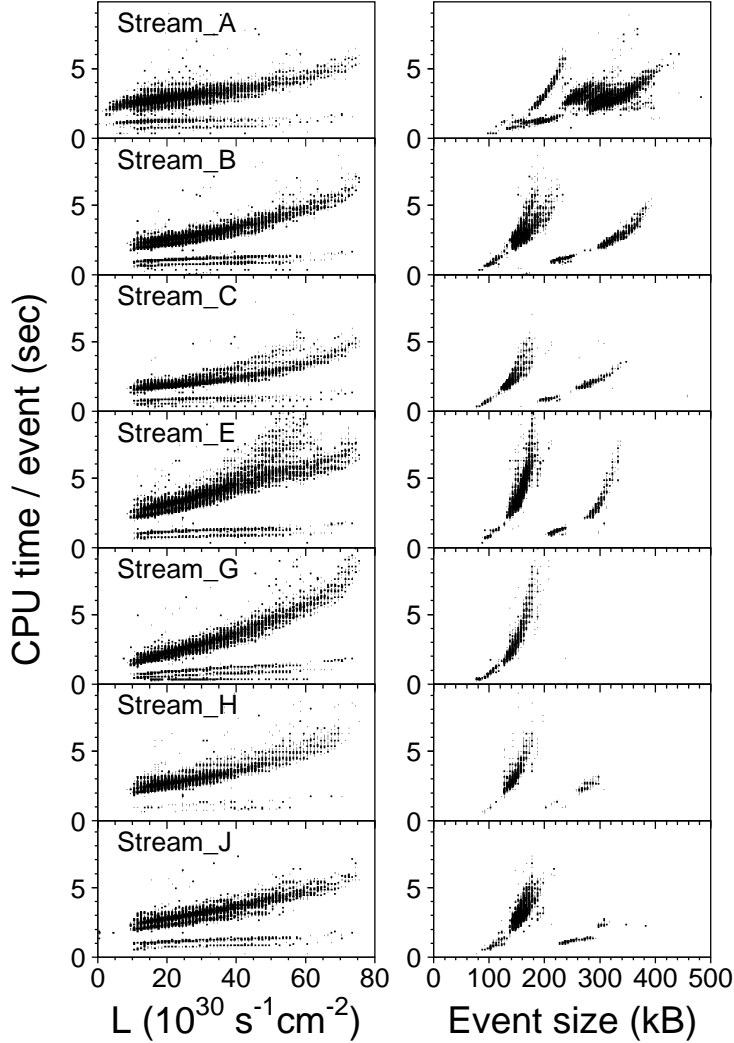,width=.75\linewidth}
  \vspace{-.4cm}
  \caption{CPU time per event versus the proton-antiproton beam 
    luminosity and event size.
    The CPU time is normalized to Pentium III 1.0 GHz.
    The separation in bands is attributed by the presence of 
    silicon detector (larger event size) and event types 
    using different reconstruction modules.
  \label{fig:CPUevt} }
\end{figure}

The event size and CPU time varies for different raw data streams.
In Fig.~\ref{fig:CPUevt} the CPU time per event
is illustrated for reconstruction of cdf software version 5.3.1.
The CPU time on a dual Pentium III 1 GHz machine varies
from 1 to 10 sec depending on the beam intensity and event size.

The input data files are staged from Enstore tapes.
The rate of staging data depends on how fast the link to Enstore 
movers is established.
Once a mover is allocated, staging a file-set of 10 GByte takes about 
20 minutes. The data transmission rate varies file by file,
the commonly observed rate is around 10 MByte/sec.

Output of concatenated files are copied to tapes.
The effectiveness in staging data to a tape is a concern because of
the limited dfarm space and output bandwidth.
A concatenation job on the output node collects files of a data-set
with close to 10 GByte at a speed that may reach the maximum IDE 
disk transfer speed of 40 MByte/sec.
It takes an average 10 minutes to copy all the files requested.
The concatenation program reads the numerous small files and writes
output that is split into into 1 GByte files. 
On a Pentium 2.6 GHz node the CPU time is about 24 minutes for 
processing 10 GByte.
The job continues by copying the output to Enstore (encp) at an average
rate of close to 20 MByte/sec.  
The encp takes about 10 minutes for writing 10 GByte.
Further delay may be caused by having more than one job accessing the
same hard disk in dfarm, or waiting to write to the same physical tape.

The output of reprocessing does not require concatenation,
(one-to-one processing with output file size of $\sim 700$ MByte).
Therefore the operation has one fewer step.
After the files are collected to output nodes, 
they are copied to Enstore tapes.
On average the stage-out takes 25 minutes for writing 
a file-set of 10 GByte to Enstore.

The tape writing is limited to one mover per data-set at a time, 
to ensure that files are written sequentially on tape.
A tape is restricted to files of the same data-set.
The instantaneous tape writing rate is 30 MByte/sec.
However, the average rate drops to below 20 MByte/sec because of 
latency in establishing connection to the mass storage system (this
includes mounting and positioning the tape establishing the end-to-end
communication).
Running only one data-set on the farm limits the capability of the farm.
Running a mix of jobs from different data-sets in parallel increases 
the through-put of the farm by increasing the output data rate.

A concatenation job (25 min for 10 GByte) spends less than half its time
accessing an Enstore mover (10 min). 
When two jobs are running for the same data-set
there is idle time while the mover is waiting for files.
The observed data transmission rate is 9 MByte/sec per data-set.
The idle time for the mover is eliminated by adding one more job 
(three concatenation for a data-set), and the data transmission rate 
increases to 15 MByte/sec.

\begin{figure}[b!]
  \centering\epsfig{file=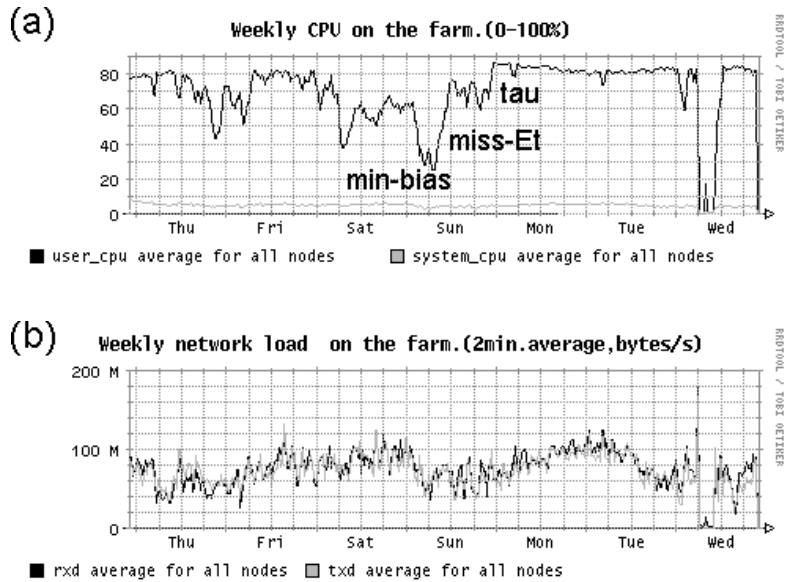,width=0.8\linewidth}
  \caption{(a) CPU load and (b) dfarm traffic 
      of the week of March 18-25, 2004.
  \label{fig:week_cpu} }
\end{figure}
  
\begin{figure}[p]
  \centering\epsfig{file=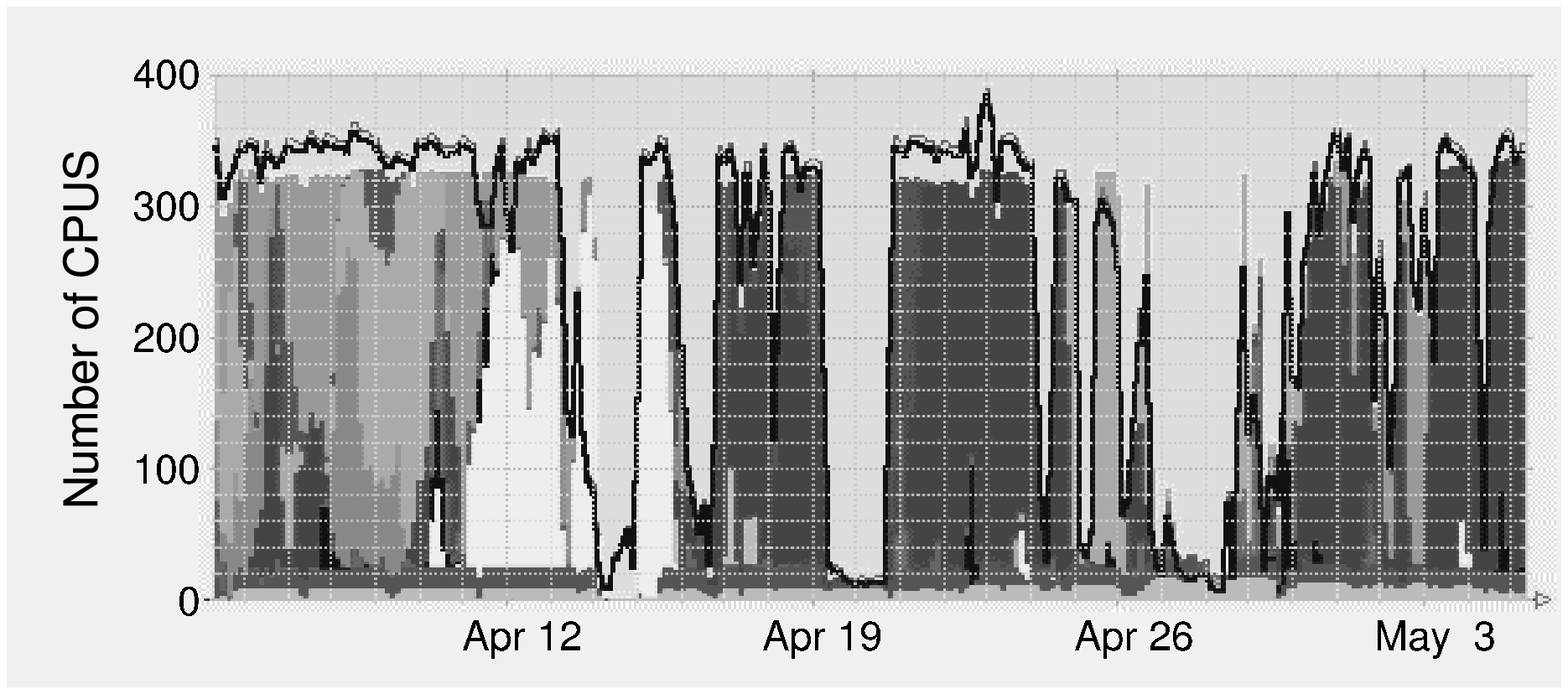,width=1.\linewidth}
  \caption{CPU usage of April 2004. Each shaded area is a 
           data-set being processed. The solid line is the sum of
           load average.
  \label{fig:month_cpu} }

\vspace{1cm}

  \hspace{2cm} \epsfig{file=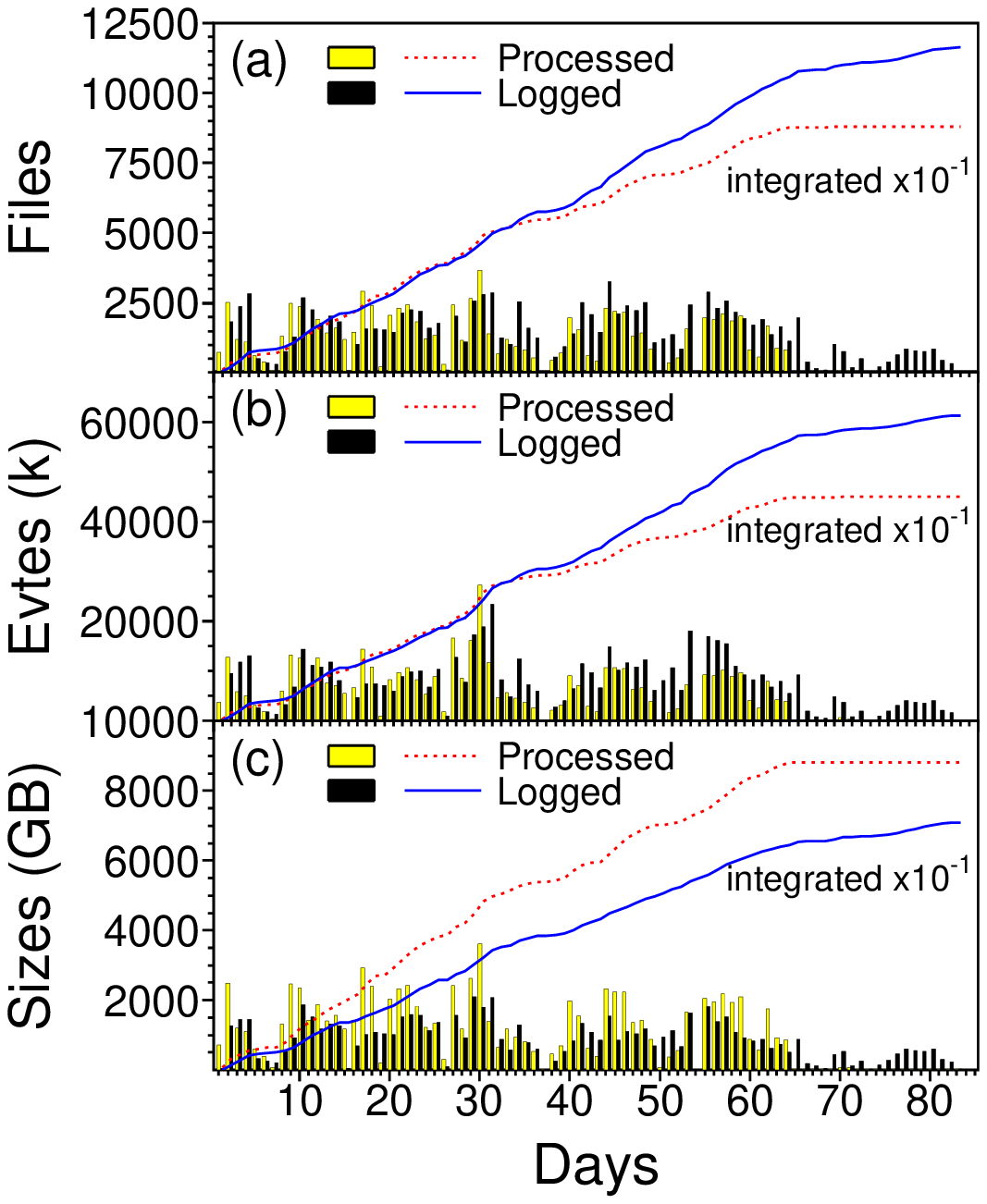,width=.70\linewidth}
  \vspace{-.5cm}
  \caption{Daily processing rates are shown in histograms for 
    (a) number of files, (b) number of events, and
    (c) data size.  The integrated rates are shown in lines.
    Compressed outputs were created for selected data-sets 
    (about a quarter of the total).
    Event size is reduced by about 30\% and thus a net reduction in 
    output storage.
  \label{fig:stat531} }
\end{figure}

The data reprocessing was performed with the revised cdf software 
version 5.3.1.
To maximize the farm efficiency the data reprocessing was performed 
on five farmlets with each farmlet processing one data-set.
The tapes were loaded one data-set at a time, therefore
farm CPU usage came in waves shared by a couple data-sets at a time.
The CPU usage for the week of March 18 is shown 
in Fig.~\ref{fig:week_cpu}.
A lag in CPU utilization was observed when the farm switched 
to a new data-set,
seen as the dips in CPU in Fig.~\ref{fig:week_cpu}.a,
because of lack of input files.
File-sets are distributed almost in sequence on a tape
The lag at the beginning of staging in a data-set is because the files 
requested are stored on the same tape, causing
all the stage-in jobs to wait for one tape.
Overall the stage-in is effective in feeding data files to dfarm.
The CPU usage varies for data-sets.
The ``minimum bias'' data-set has smaller file sizes and
the CPU per event is about 40\% less than the average.
When this data-set was processed, the stage-in rate was not 
able to keep up with the CPU consumption.

The production farm operation is efficient.
The CPU usage for the month of April 2004 is shown 
in Fig.~\ref{fig:month_cpu}.
Each shaded area seen is one data-set being processed.
The output data logging rate is shown in Fig.~\ref{fig:stat531} for
the number of files, number of events, and total file size written 
to Enstore tapes.
Compressed outputs were also created for selected data-sets.
Therefore the total events in output was increased by about 25\%.
The event size was reduced and resulted to a net reduction in 
storage by about 20\%.
On average we had a through-put of over 2 TByte (10 million events) 
per day to the Enstore storage.
The data logging lasted two extra weeks for a large B physics data-set
that accounted for about 20\% of the total CDF data. 
It was the latest data-set processed and the tape logging rate 
was saturated at about 800 GByte per day.

\begin{figure}[b!]
  \centering\epsfig{file=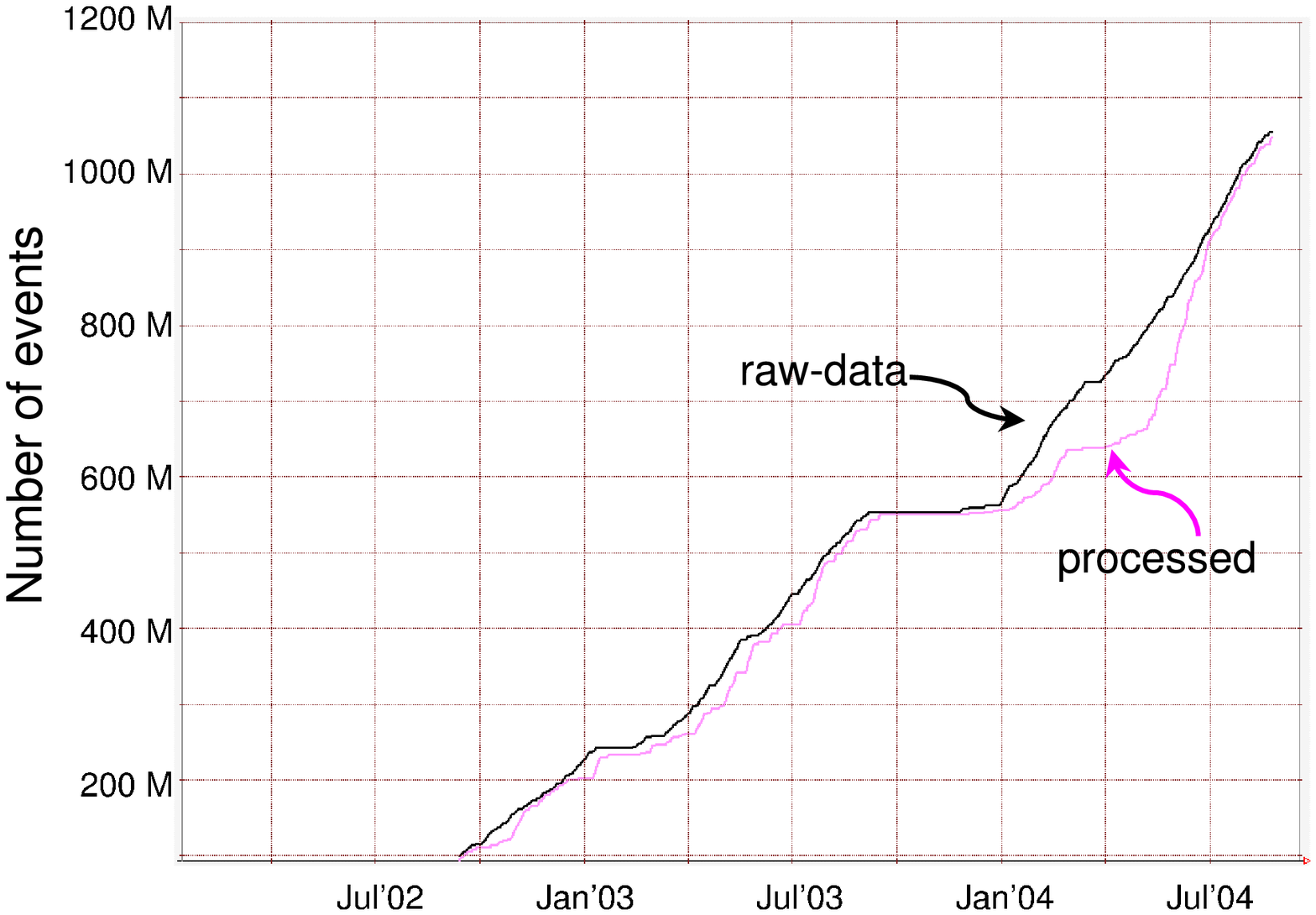,width=0.75\linewidth}
  \caption{The cumulative raw-data and production data rate
           for the period processed with cdf software version 5.3.1.
  \label{fig:dual-line3} }
\vspace{-.5cm}
\end{figure}

\subsection{Processing recently acquired data }

The production farm uses cron jobs to check the online database 
for newly acquired data.
Timely processing is critical for detector monitoring.
The express Stream-A processing is used for monitoring data quality 
and beam-line calibrations are performed using Stream-G data.
Data-sets of Stream-B and Stream-G are used for additional calibrations.
Full data processing is then carried
out after final calibration constants are available.

The load on the farm for new data, at a logging rate of 10 pb$^{-1}$ 
per week, is less than half of the CPU capacity.
The raw-data volume collected and processed are shown in 
Fig.~\ref{fig:dual-line3}.
In February 2004 one of the major detector components was unstable.
Raw data processing was held except for detector studies.
Meanwhile the farm was put in use for 5.3.1 reprocessing.

Raw-data processing resumed in early May 2004.
Data collected after February 2004 were processed 
with preliminary calibrations.
Later it was reprocessed with refined calibrations.
The reprocessing was started in October 2004.

\section{Conclusion}

The CDF production PC farms have been successfully prototyped, 
installed, integrated, commissioned and operated for many years. 
They have been successful in providing the computing processing capacity
required for the CDF experiment in Run II.  The success of this system 
has in turn enabled successful analysis of the wealth of new data being 
collected by the CDF experiment at Fermilab.  The system has been 
modified and enhanced during the years of its operation to adjust 
to new requirements and to enable new capabilities.  

The system will continue to be modified in the future to continue to 
serve the CDF collaboration as required.  Some of the modifications 
are simple upgrades of components with more capable (faster, more 
capacious) replacements.  Other modifications will affect the 
architecture of the system and quite likely will embrace distributed 
processing and the grid in some way.  These developments will allow 
CDF to continue to process and analyze data through the end of the 
life of the experiment.

{}

\end{document}